\RequirePackage{fix-cm}
\RequirePackage{fixltx2e}
\documentclass[reqno,intlimits]{amsproc}
\usepackage{amsthm,amssymb}
\usepackage{hyperref}

\usepackage[citation-order,nobysame]{amsrefs}
\usepackage{xyzbib}

\numberwithin{equation}{section}
\let\cite=\cites
\newcommand{\rmd}{\mathrm{d}}
\newcommand{\rmi}{\mathrm{i}}
\newcommand{\rme}{\mathrm{e}}

\newcommand{\kf}{k_\text{F}}

\newcommand{\cc}{c}

\newcommand{\vac}{\Omega_\Uparrow}
\newcommand{\up}{\uparrow}
\newcommand{\dn}{\downarrow}
\newcommand{\wt}{\widetilde}

\newcommand\bra[1]{\left\langle #1 \right\rvert}
\newcommand\ket[1]{\left\lvert #1 \right\rangle}
\newcommand\bracket[2]{\left\langle #1 \vert #2 \right\rangle}

\DeclareMathOperator{\arccot}{arccot}
\DeclareMathOperator{\sgn}{sgn}

\begin{document}

\title{Impurity Green's function of a one-dimensional Fermi gas}

\author{Oleksandr Gamayun}
\address{Lancaster University, Physics Department, Lancaster, LA1 4YB, UK
and Bogoluibov Institute for Theoretical Physics, Metrolohichna 14-b, 
Kiev, 03680,Ukraine}

\author{Andrei G. Pronko}
\address{V. A. Steklov Mathematical Institute,
Fontanka 27, St.~Petersburg, 191023, Russia}

\author{Mikhail B. Zvonarev}
\address{Univ Paris-Sud, Laboratoire LPTMS, UMR8626, Orsay, F-91405, France
and CNRS, Orsay, F-91405, France}

\begin{abstract}
We consider a one-dimensional gas of spin-1/2 fermions
interacting through $\delta$-function repulsive potential of an
arbitrary strength. For the case of all fermions but one having spin
up, we calculate time-dependent two-point correlation function of the
spin-down fermion. This impurity Green's function is represented in the 
thermodynamic limit as an integral of Fredholm determinants of 
integrable linear integral operators.

\end{abstract}

\maketitle
\section{Introduction}

Consider a one-dimensional system of quantum particles
driven by the secondary quantized Hamiltonian
\begin{equation}\label{Ham}
H=\int_0^L\rmd x\,
\left(\sum_{s=\up,\dn}
\partial_x\psi_s^\dagger(x)\partial_x\psi_s(x)
+2\cc\,\psi_\up^\dagger(x)\psi_\up(x)\psi_\dn^\dagger(x) \psi_\dn(x)\right),
\end{equation}
where $\cc$ is a coupling constant, $\cc>0$. 
The one-dimensional quantum fields $\psi_s(x)$ and $\psi_s^\dagger(x)$ ($s=\up,\dn$) satisfy canonical anti-commutation relations
\begin{equation}\label{anticomm}
\psi_s(x)\psi_{s'}^\dagger(x')+ \psi_{s'}^\dagger(x') \psi_s(x)
= \delta_{ss'}\delta(x-x'),
\end{equation}
with all other anticommutators vanishing.
The impurity Green's function is the two-point correlation function
\begin{equation}\label{Gdef}
G_\downarrow(x,t)=\bra{\vac}
\rme^{\rmi t H} \psi_\dn(x) \rme^{-\rmi t H}\psi_\dn^\dagger(0)
\ket{\vac},
\end{equation}
where $\ket{\vac}$ is the normalized ground state of $N$ spin up fermions, 
$\bracket{\vac}{\vac}=1$. 
The function \eqref{Gdef} describes a propagation of an impurity
(spin-down fermion) in a gas of free (spin-up) fermions. The impurity
interacts with free fermions through  a $\delta$-function potential of
strength $2c$.

Eigenfunctions and spectrum of the second-quantized Hamiltonian \eqref{Ham}
in the sector with $N$ spin-up and one spin-down fermions were found by
McGiure \cite{McG-65}. Representations of eigenfunctions in terms of determinants
resembling Slater determinant for free fermions were discussed 
by Edwards \cite{E-90},  Castella and Zotos \cite{CZ-93}, 
and Recher and Kohler \cite{RK-12}. A
solution to the eigenstates problem for the Hamiltonian \eqref{Ham} in a
sector with arbitrary number of spin-up and spin-down fermions was
given by C.\,N.~Yang \cite{Y-67} and Gaudin, see \cite{Ga-14} and references therein.

In this paper we calculate the impurity Green's function \eqref{Gdef}.
By exploiting the determinant structure of the eigenfunctions
of the problem we employ the technique, known previously in the case
of infinite coupling, which leads to determinant representations for
form factors and correlation functions. Namely, for the system in a
finite volume $L$ and for finite number $N$ of spin-up fermions, we
represent the impurity Green's function \eqref{Gdef} in terms of
determinants of $N\times N$ matrices. Taking the thermodynamic limit,
defined as the limit $N,L\to \infty$ with the ratio $N/L$ kept
constant, we get the Green's function \eqref{Gdef} expressed in terms
of Fredholm determinants of integrable linear integral operators, as
shown by formula \eqref{mainresult}.

Representations in terms of Fredholm determinants for correlation functions
were introduced, on the example of density matrix of the Tonks-Gerardeau gas
(impenetrable Bose gas), by Schultz \cite{S-63} and
Lenard \cite{L-64,L-66}. Subsequently, it was
realized that the representations of this kind plays an important role in
description of correlation functions of one-dimensional
quantum solvable models in terms
of classical integrable systems \cite{SMJ-79,JMMS-80}.
The role of integrable integral operators has been realized in
\cite{IIKS-90}. The whole approach makes it possible to construct
asymptotic expansions of correlation functions, e.g., for long times
and large distances using the method of matrix
Riemann-Hilbert problem \cite{KBI-93,DIZ-97}; for a recent progress, see
\cite{KKMST-09,S-10,K-11}.

At the infinite coupling the 
general, depending on temperature, time, external magnetic field
and chemical potential, two-point correlation functions of the
two-component Fermi and Bose gases were computed in terms of Fredholm determinants
in \cite{IP-97,IP-98}. Similar results in the context of the Hubbard model
and ladder spin (a spin-1/2 Bose-Hubbard) model at the
infinite coupling were given in \cite{IPA-98,AIP-01}.
Various asymptotic results about these correlation functions, using
the Reimann-Hilbert problem, have been obtained in
\cite{GIKP-98,GIk-98,ChZ-04a,ChZ-04b}. Applications of these
results to description of physical phenomena are
discussed in \cite{GK-99,ChSZ-05,ChZ-08,ZChG-09}.

Recently, a number of new phenomena were predicted for the dynamics
and relaxation of the mobile impurity injected into a free Fermi gas
\cite{L-09,MZD-12,SGK-12,SKGL-12,KMGZD-13,BChGL-14,G-14,GLCh-14,KSG-14}. 
We expect that our findings will pave a way towards quantitative 
description of those phenomena.

\section{Bethe Ansatz and the impurity problem}

The basis in the Fock space of the model is constructed by acting with
operators $\psi^\dagger_s(x)$ onto the pseudovacuum $\ket{0}$, defined as
\begin{equation}
\psi_s(x)\ket{0}=0,\qquad
\bra{0}\psi_s^\dagger(x)=0,\qquad
\bracket{0}{0}=1.
\end{equation}
We say that a state belongs to the sector $(N,M)$ of the Fock space if
it contains $N-M$ particles of spin up and $M$ particles
of spin down, so that the total number of particles is $N$.
The number of particles of each type being conserved separately, an
eigenstate of the Hamiltonian \eqref{Ham} can be obtained as a linear superposition
of the basis states from the same sector. In the sector $(N,M)$ the
eigenstates are enumerated by two sets, $\{k\}=k_1,\ldots,k_N$ and
$\{\lambda\}=\lambda_1,\ldots,\lambda_M$, of unequal, in each set
separately, numbers, called quasi-momenta.

In the sector $(N,M)$ the eigenstates can be written in the form
\begin{equation}\label{PsiNM}
|\Psi_{N,M}(\{k\};\{\lambda\})\rangle
=\int_{[0,L]^{N}}\!\rmd^N x
\sum_{\{s\}}^{}
\Psi_{N,M}^{\{s\}}(\{k\},\{\lambda\}|\{x\})\,
\psi^\dagger_{s_1}(x_1)\cdots \psi^\dagger_{s_N}(x_N) \ket{0},
\end{equation}
where $\{x\}:=x_1,\dots,x_N$ and the wave function
$\Psi_{N,M}^{\{s\}}(\{k\},\{\lambda\}|\{x\})$
is not equal to zero only if $M$ elements are equal to $\dn$ and
$N-M$ elements are equal to $\up$ in the set $\{s\}:=s_1,\dots,s_N$.
The quasi-momenta are solutions of the nested Bethe Ansatz equations:
\begin{align}\label{nest}
\rme^{\rmi k_jL}
&=\prod_{a=1}^{M}\frac{k_j-\lambda_a+\rmi\cc/2}{k_j-\lambda_a-\rmi\cc/2},
\qquad j=1,\dots,N,
\\  \label{nest2}
\prod_{j=1}^{N}\frac{k_j-\lambda_a+\rmi\cc/2}{k_j-\lambda_a-\rmi\cc/2}
&=\prod_{b=1}^{M}
\frac{\lambda_b-\lambda_a+\rmi\cc}{\lambda_b-\lambda_a-\rmi\cc},\qquad a=1,\dots,M.
\end{align}
The eigenvalue (energy) of the Hamiltonian \eqref{Ham} in the
sector $(N,M)$ is
\begin{equation}
E_{N,M}(\{k\},\{\lambda\})=\sum_{j=1}^{N}k_j^2.
\end{equation}
For what follows we
need only the eigenstates belonging to the sectors $(N,0)$ and
$(N+1,1)$. The former are given simply by the Slater determinant wave function
while the latter can be constructed using a connection to the problem of
mobile impurity interacting with the gas of $N$ free
fermions.

In the sector $(N,0)$ the second set of quasi-momenta is empty, $\{\lambda\}=\emptyset$,
and the Bethe equations for the elements of the first set
are given by \eqref{nest} with the right-hand side equal to one.
For the sake of further convenience in calculations, we denote
the quasi-momenta of the first set as
$\{q\}=q_1,\dots,q_N$. The eigenstate \eqref{PsiNM} in this case reads
\begin{equation}
\ket{\Psi_{N,0}(\{q\})}=\int_{[0,L]^N}\rmd^N x\,
\Psi_{N,0}^{\up\dots\up}(\{q\}|\{x\})\,\psi_\up^\dagger(x_1)\cdots\psi_\up^\dagger(x_N)\ket{0},
\end{equation}
where the wave function is
\begin{equation}\label{PsiN0}
\Psi^{\up\dots\up}_{N,0}(\{q\}|\{x\})=
\frac{1}{N!L^{N/2}}\det_{1\leq j,l\leq N} \left(\rme^{\rmi q_j x_l}\right).
\end{equation}
The quasi-momenta are quantized as
\begin{equation}\label{qj}
q_j=\frac{2\pi}{L} m_j,\qquad
m_j\in\mathbb{Z},
\end{equation}
and in order that the wave function does not vanish, they must all be distinct;
the eigenstates in the considered sector are labeled in a unique way
by sets $\{q\}$ in which $q_1<q_2<\dots<q_N$.
The ground state in the sector $(N,0)$ is the state 
$\ket{\vac}=\ket{\Psi_{N,0}(\{q\})}$, in which 
$-\kf\leq q_1<\dots<q_N\leq \kf$ for $N$ odd,
and $-\kf< q_1<\dots<q_N\leq \kf$ or $-\kf\leq q_1<\dots<q_N<\kf$ for $N$ even, 
where $\kf=\pi N/L$.

Now we turn to construction of the eigenstates in the sector $(N+1,1)$.
According to \eqref{PsiNM} they have the form
\begin{multline}\label{PsiN11}
\ket{\Psi_{N+1,1}(\{k\},\lambda)}
=\int_{[0,L]^{N+1}}
\rmd^{N+1} x\,
\sum_{j=1}^{N+1}\Psi_{N+1,1}^{\Uparrow_j}(\{k\},\lambda|\{x\})
\\ \times
\psi_\up^\dagger(x_1)\cdots\psi_\up^\dagger(x_{j-1})\psi_\dn^\dagger(x_j)
\psi_\up^\dagger(x_{j+1})\cdots\psi_\up^\dagger(x_{N+1})
\ket{0},
\end{multline}
where $\Uparrow_j$ denotes the set of $N+1$ spins with all but one spins up,
$\Uparrow_j=\up\dots\up\dn\up\dots\up$, with the down spin standing on the
$j$th position. 
We relabel arguments $x_1,\dots,x_{N+1}$ in \eqref{PsiN11} in
such a way that the field operator $\psi^\dagger_\dn$ always depends
on $x_{N+1}$:
\begin{multline}
\ket{\Psi_{N+1,1}(\{k\},\lambda)}
=\int_{[0,L]^{N+1}}
\rmd^{N+1} x\,
\wt\Psi_{N+1,1}(\{k\},\lambda|\{x\})
\\ \times
\psi_\up^\dagger(x_1)\cdots\psi_\up^\dagger(x_{N})\psi_\dn^\dagger(x_{N+1})
\ket{0}.
\end{multline}
The wave function
$\wt\Psi_{N+1,1}(\{x\})$ (to ease notations,
we do not show below the dependence of the wave-functions on quasi-momenta
explicitly whenever possible) 
is expressed in terms of those entering \eqref{PsiN11} as follows
\begin{equation}
\wt\Psi_{N+1,1}(\{x\})
=\sum_{j=1}^{N+1}(-1)^{N-j}
\Psi_{N+1,1}^{\Uparrow_j}(x_1,\dots,x_{j-1},x_{N+1},x_{j+1},\dots,x_{N}),
\end{equation}
and it must be an eigenfunction of the first quantized Hamiltonian
\begin{equation}\label{HN11}
\mathcal{H}_{N+1,1}=
-\sum_{j=1}^{N+1}\frac{\partial^2}{\partial x_j^2}
+2\cc \sum_{j=1}^{N} \delta(x_i-x_{N+1}),
\end{equation}
where $x_1,\dots,x_{N+1}\in[0,L]$, with
the periodic boundary conditions.

A way to diagonalize the Hamiltonian
\eqref{HN11} used in \cite{E-90,CZ-93} is to represent 
it first in the reference
frame of the spin down particle, introducing the coordinates
\begin{equation}
y_j=x_j-x_{N+1},\qquad j=1,\dots,N.
\end{equation}
Hence, $x_j=y_j+x_{N+1}$, $j=1,\dots,N$, and therefore
\begin{equation}
\frac{\partial}{\partial x_j}=\frac{\partial}{\partial y_j},\quad
j=1,\dots,N,\qquad
\frac{\partial}{\partial x_{N+1}}
=\rmi \mathcal{P}_{N+1}- \sum_{j=1}^{N}\frac{\partial}{\partial y_j},
\end{equation}
where $\mathcal{P}_{N+1}$ is the first quantized
total momentum operator of $N+1$ particles,
\begin{equation}
\mathcal{P}_{N+1}=\sum_{j=1}^{N+1}\frac{1}{\rmi}\frac{\partial}{\partial x_j}.
\end{equation}
Since $\mathcal{P}_{N+1}$ is an integral of motion,
$[\mathcal{P}_{N+1},\mathcal{H}_{N+1,1}]=0$,
it can be replaced by its eigenvalues $P$ when acting on the eigenfunctions,
\begin{equation}\label{totalP}
P=\sum_{j=1}^{N+1}k_j=\frac{2\pi}{L} n,\qquad n\in\mathbb{Z}.	
\end{equation}
The Hamiltonian in the new coordinates reads
\begin{equation}\label{Himp}
\mathcal{H}_{N+1,1}=
-\sum_{j=1}^{N}\frac{\partial^2}{\partial y_j^2}
+\Bigg(P+\rmi\sum_{j=1}^{N}\frac{\partial}{\partial y_j} \Bigg)^2
+2\cc \sum_{j=1}^{N} \delta(y_i).
\end{equation}
Noticing that this in fact an $N$-particle Hamiltonian and denoting its
eigenfunctions as $\Phi_N(\{y\})$, where $\{y\}:=y_1,\dots, y_N$, 
we have the following
relation between the wave functions in the two reference frames \cite{CZ-93}:
\begin{equation}\label{twoframes}
\wt\Psi_{N+1,1}(\{x\})=\rme^{\rmi P x_{N+1}}
\Phi_N(x_1-x_{N+1},\dots,x_N-x_{N+1}).
\end{equation}
The function $\Phi_N(\{y\})=\Phi_N(\{k\},\lambda|\{y\})$ must be periodic function
in each coordinate, and totally antisymmetric with respect to their permutations.

We construct the function $\Phi_N(\{y\})$ restricting to the domain 
$y_1,\dots, y_N\in[0,L]$. The periodicity implies that 
\begin{equation}\label{period}
\Phi_N(\{y\})|_{y_j=0}=\Phi_N(\{y\})|_{y_j=L},\qquad
j=1,\dots,N,
\end{equation}
and that the first derivatives, due to the
$\delta$-function potential in \eqref{Himp}, must satisfy the conditions
\begin{equation}\label{jump}
\frac{\partial}{\partial y_j}\Phi_N(\{y\})\bigg|_{y_j=0}-
\frac{\partial}{\partial y_j}\Phi_N(\{y\})\bigg|_{y_j=L}=\cc
\Phi_N(\{y\})|_{y_j=0},\qquad
j=1,\dots,N.
\end{equation}
The antisymmetry implies that $\Phi_N(\{y\})$ is
a determinant
\begin{equation}\label{PhiDet}
\Phi_N(\{y\})=C_{\Phi_N}\det_{1\leq j,l\leq N}(\phi_j(y_l)).
\end{equation}
Here $C_{\Phi_N}$ is a normalization factor to be found later, and the
functions $\phi_j(y)$ can be represented as a superpositions of at least $N+1$
plane waves:
\begin{equation}
\phi_j(y)=\sum_{l=1}^{N+1} a_{jl} \rme^{\rmi k_l y}.
\end{equation}
The condition \eqref{period} implies that
\begin{equation}\label{shortcond}
\sum_{l=1}^{N+1}a_{jl}(1-\rme^{\rmi k_l L})=0,\qquad
j=1,\dots,N,
\end{equation}
and \eqref{jump} that
\begin{equation}\label{longcond}
\sum_{l=1}^{N+1}a_{jl}\left(\rmi k_l(1-\rme^{\rmi k_l L})-\cc\right)=0,\qquad
j=1,\dots,N.
\end{equation}
Subtracting \eqref{shortcond} from \eqref{longcond} with the factor
$\rmi \lambda-\cc/2$, where $\lambda$ is the quasi-momentum of
the auxiliary Bethe Ansatz problem, and
requiring that all coefficients of the sum in the resulting equation
vanish, we reproduce the first set of Bethe Ansatz equations \eqref{nest}
of the sector $(N+1,1)$:
\begin{equation}\label{BAEkv}
\rme^{\rmi k_l L}
=\frac{k_l-\lambda+\rmi \cc/2}{k_l-\lambda-\rmi\cc/2},\qquad l=1,\dots,N+1.
\end{equation}
Recalling the quantization condition for the total momentum \eqref{totalP},
we also have the equation for $\lambda$:
\begin{equation}\label{BAEvk}
\prod_{l=1}^{N+1}\frac{k_l-\lambda+\rmi \cc/2}{k_l-\lambda-\rmi\cc/2}=1.
\end{equation}
Obviously, this equation is exactly the second set of the Bethe Ansatz equations
of the sector $(N+1,1)$, see \eqref{nest2}.

The Bethe Ansatz equations \eqref{BAEkv} can also be written in the form
\begin{equation}\label{BAcot}
\cot \frac{k_j L}{2}= \frac{2(k_j-\lambda)}{\cc},\qquad
j=1,\dots,N+1.
\end{equation}
Introduce the notation
\begin{equation}\label{deltajs}
\alpha_j=-\arccot\frac{2(k_j-\lambda)}{\cc}, \qquad j=1,\dots,N+1.
\end{equation}
We use the convention that $\arccot x\in [0,\pi]$, $x\in \mathbb{R}$.
The quantization conditions for momenta $\{k\}$ are
\begin{equation}
k_j=\frac{2\pi}{L}n_j-\frac{2}{L}\alpha_j,\qquad
n_j\in\mathbb{Z}.
\end{equation}
Equation \eqref{BAEvk} has the form
\begin{equation}
\frac{R_N(\lambda)}{Q_{N+1}(\lambda)}=0,\qquad
\end{equation}
where $R_N(\lambda)$ and $Q_{N+1}(\lambda)$ are polynomials in $\lambda$ of
the degrees shown in the subscripts,
\begin{equation}
R_N(\lambda)=Q_{N+1}(\lambda-\rmi\cc)-Q_{N+1}(\lambda),\qquad
Q_{N+1}(\lambda)=\prod_{l=1}^{N+1}\left(k_l-\lambda-\frac{\rmi\cc}{2}\right).	
\end{equation}
The quantization condition for the
quasi-momentum $\lambda$, which takes $N+1$ values, follows from the relation
\begin{equation}\label{sumdelta}
\sum_{j=1}^{N+1}\alpha_j=-\pi m,
\qquad m=0,1,\dots,N.
\end{equation}
The value $m=0$ corresponds to $\lambda=-\infty$; the values
$m=1,\dots,N$ correspond to the roots of the polynomial $R_{N}(\lambda)$.

To finalize the construction of the wave function $\Phi_N(\{y\})$,
we have to satisfy the
conditions \eqref{shortcond}. It can be easily seen that this can be done
by choosing
\begin{equation}
\phi_j(y)=\chi_{j}(y)-\chi_{N+1}(y),
\qquad j=1,\dots,N,
\end{equation}
where
\begin{equation}
\chi_l(y)=\frac{\rme^{\rmi k_l y+\rmi\alpha_l}}{\sin\alpha_l}
=-\frac{2}{\cc}\left(k_l-\lambda-\frac{\rmi\cc}{2}\right) \rme^{\rmi k_l y},\qquad
l=1,\dots,N+1.
\end{equation}
With our choice it is obvious that the wave function $\Phi_N(\{y\})$
can also be represented as the following determinant of an $(N+1)\times (N+1)$ matrix:
\begin{equation}\label{PhiBigDet}
\Phi_N(\{y\})=
C_{\Phi_N}
\begin{vmatrix}
\chi_1 (y_1) &\dots & \chi_1 (y_N) & 1\\
\vdots & \ddots &\vdots &\vdots\\
\chi_{N+1} (y_1) &\dots & \chi_{N+1} (y_N) & 1\\
\end{vmatrix}.
\end{equation}

Recalling relation \eqref{twoframes} one can readily obtain 
the corresponding expression for the wave function $\wt\Psi_{N+1,1}(\{x\})$. 
The only point need to be taken into account is that the formulas above 
have to be extended on the values $y_1,\dots,y_N\in [-L,L]$.
Due to the periodicity, for $y\in[-L,0]$ we can set $\chi_l(y)=\chi_l(L+y)$ 
and using the Bethe Ansatz equations \eqref{BAEkv}, we have
\begin{equation}\label{chiext}
\chi_l(y)=-\frac{2}{c}
\left(k_j-\lambda-\frac{\rmi\cc}{2}\sgn y\right)\rme^{\rmi k_l y},\qquad y\in [-L,L],
\end{equation}
where $\sgn y$ is the signum function. 
It is worth to mention that the resulting 
expression for function $\wt\Psi_{N+1,1}(\{x\})$ which follows 
from \eqref{twoframes}, \eqref{PhiBigDet}
and  \eqref{chiext} 
coincides with the expression for the wave function 
proposed in \cite{RK-12}.

To satisfy the normalization condition
$\bracket{\Psi_{N+1,1}(\{k\},\lambda)}{\Psi_{N+1,1}(\{k\},\lambda)}=1$ the
constant $C_{\Phi_N}$ has to be chosen such that
\begin{equation}\label{CPhiN}
|C_{\Phi_N}|^2=\frac{1}{(N!)^2 L^{N+1}}
\left(\sum_{l=1}^{N+1}
\prod_{\substack{j=1\\ j \ne l}}^{N+1}
\left(\frac{1}{{\sin}^2\alpha_j}+\frac{4}{L\cc}\right)
\right)^{-1}.
\end{equation}
We prove \eqref{CPhiN} in the next section.

\section{Finite volume calculations}

In this section we perform various calculations with the
wave functions obtained in the previous section.
We first prove formula \eqref{CPhiN}. We have
\begin{multline}\label{scalar}
\bracket{\Psi_{N+1,1}(\{k\},\lambda)}{\Psi_{N+1,1}(\{k\},\lambda)}
=N! L \int_{[0,L]^N}\rmd^N y\, |\Phi_N(\{y\})|^2
\\
=|C_{\Phi_N}|^2 (N!)^2 L^{N+1}
\det_{1\leq j,l \leq N}
\left(\frac{1}{L}\int_{0}^L\rmd y\, \overline{\phi_j(y)}\phi_l(y) \right).
\end{multline}
To evaluate the entries of the matrix here, we use
the following relation, valid for $k_j\ne k_l$,
\begin{equation}
\int_{0}^{L}\rmd y\, \rme^{\rmi(k_l-k_j)y}
= \frac{\rme^{\rmi(k_l-k_j)L}-1}{\rmi(k_l-k_j)}
= - \frac{4\sin\alpha_j\sin\alpha_l}{\cc}\rme^{\rmi\alpha_j-\rmi\alpha_l},
\end{equation}
where the second equality follows by considering a
difference of two Bethe Anzatz equations \eqref{BAcot}. For the off-diagonal entries,
we obtain
\begin{equation}
\frac{1}{L}\int_{0}^L\rmd y\,
\overline{\phi_j(y)}\phi_l(y) = \frac{1}{{\sin}^2\alpha_{N+1}}-\frac{4}{L\cc}
\qquad
(j\ne l),
\end{equation}
and for the diagonal ones, we obtain
\begin{equation}
\frac{1}{L}\int_{0}^L\rmd y\, |\phi_j(y)|^2
= \frac{1}{{\sin}^2\alpha_{j}}
+\frac{1}{{\sin}^2\alpha_{N+1}}-\frac{8}{L\cc}.
\end{equation}
Denoting $u_j=1/{\sin}^2\alpha_j-4/L\cc$,
$j=1,\dots,N+1$, we see that the $N\times N$
matrix in the determinant in \eqref{scalar}
is the sum of a diagonal matrix, with entries $u_j\delta_{jl}$,
and of a matrix of rank one, with all entries equal to $u_{N+1}$.  Hence,
\begin{multline}
\det_{1\leq j,l \leq N}
\left(\int_{0}^L\rmd y\, \overline{\phi_j(y)}\phi_l(y) \right)=
\det_{1\leq j,l \leq N}\left(u_j\delta_{jl}+u_{N+1}\right)
\\
=u_1\cdots u_N \left(1+\left(u_1^{-1}+\cdots+u_N^{-1}\right)u_{N+1}\right)
=\sum_{j=1}^{N+1}u_j^{-1}\prod_{j=1}^{N+1}u_j,
\end{multline}
and formula \eqref{CPhiN} follows.

Next we consider the form-factors of the
operator $\psi_\dn(x)$ relevant to the Green's function
\eqref{Gdef}. Specifically, these are its matrix elements between the states
belonging to the sectors $(N,0)$ and $(N+1,1)$. We have
\begin{multline}
\bra{\Psi_{N,0}(\{q\})}\rme^{\rmi t H}
\psi_\dn(x)\rme^{-\rmi t H}\ket{\Psi_{N+1,1}(\{k\},\lambda)}
\\
=
\bra{\Psi_{N,0}(\{q\})}
\psi_\dn(0)\ket{\Psi_{N+1,1}(\{k\},\lambda)}
\exp\left\{\rmi\sum_{j=1}^{N}\tau(q_j)
-\rmi\sum_{j=1}^{N+1}\tau(k_j)\right\},
\end{multline}
where $\tau(q)=t q^2 -x q$.
Using expressions \eqref{PsiN0} and \eqref{PhiBigDet}
for the wave functions involved, for the form-factor
we obtain the following representation:
\begin{multline}\label{FFD}
\bra{\Psi_{N,0}(\{q\})}\psi_\dn(0)\ket{\Psi_{N+1,1}(\{k\},\lambda)}
\\
=N! \int_{[0,L]^N}\rmd^N x\,
\overline{\Psi_{N,0}^{\up\dots\up}(\{x\})}\Phi_N(\{x\})
=\frac{N! C_{\Phi_N}}{L^{N/2}} F_N(\{q\}|\{k\}),
\end{multline}
where
\begin{equation}
F_N(\{q\}|\{k\}) = \det_{1\leq j,l \leq N} \left(
\int_{0}^L \rmd x\, \rme^{-\rmi q_j x}\phi_l(x)\right).
\end{equation}
Taking into account that
$\rme^{\rmi q_j L}=1$ and $\rme^{\rmi k_l L}=\rme^{-2\rmi\alpha_l}$,
the integrals are evaluated as follows:
\begin{equation}
\int_{0}^L \rmd x\, \rme^{-\rmi q_j x} \chi_l(x)
=\frac{\rme^{\rmi \alpha_l}}{\sin\alpha_l}
\int_0^L\rmd x\, \rme^{\rmi(k_l-q_j)x}
=\frac{\rme^{\rmi \alpha_j}}{\sin\alpha_j}
\frac{\rme^{\rmi(k_l-q_j)L}-1}{\rmi(k_j-q_l)}
=\frac{2}{q_j-k_l}.
\end{equation}
Hence,
\begin{equation}\label{FNN}
F_N(\{q\}|\{k\})=2^N
\det_{1\leq j,l\leq N}
\left(\frac{1}{q_j-k_l}-\frac{1}{q_j-k_{N+1}}\right).
\end{equation}
We also note that the function
$F_N(\{q\}|\{k\})$ can be written as a
determinant of an $(N+1)\times(N+1)$ matrix,
\begin{equation}\label{FN1}
F_N(\{q\}|\{k\})=
2^N
\begin{vmatrix}
(q_1-k_1)^{-1} &\dots & (q_1-k_{N+1})^{-1}\\
\vdots & \ddots & \vdots \\
(q_N-k_1)^{-1} &\dots & (q_N-k_{N+1})^{-1}\\
1 & \dots  & 1\\
\end{vmatrix},
\end{equation}
which can be obtained when using \eqref{PhiBigDet} instead of \eqref{PhiDet}.

We now consider the impurity Green's function on the finite lattice,
defined as a diagonal matrix element of the
two-point field operator in the sector $(N,0)$,
\begin{equation}
G_{\dn,N}(x,t|\{q\})=\bra{\Psi_{N,0}(\{q\})}\rme^{\rmi t H}
\psi_\dn(x)\rme^{-\rmi t H}\psi_\dn^\dagger(0)\ket{\Psi_{N,0}(\{q\})}.
\end{equation}
This function can be written as the sum over
all states (i.e., all distinct solutions of the
Bethe Ansatz equations) in the sector $(N+1,1)$:
\begin{multline}
G_{\dn,N}(x,t|\{q\})=\sum_{\{k\},\lambda}
\big|\bra{\Psi_{N,0}(\{q\})}
\psi_\dn(0)\ket{\Psi_{N+1,1}(\{k\},\lambda)}\big|^2
\\ \times
\exp\left\{\rmi\sum_{j=1}^{N}\tau(q_j)
-\rmi\sum_{j=1}^{N+1}\tau(k_j)\right\}.
\end{multline}
Using \eqref{FFD} and taking into account \eqref{CPhiN},
we have
\begin{multline}\label{Gsum}
G_{\dn,N}(x,t|\{q\})
= \frac{1}{L^{2N+1}}\sum_{\{k\},\lambda}
\frac{u_1^{-1}\cdots u_{N+1}^{-1}}{u_1^{-1}+\dots+u_{N+1}^{-1}}
\\ \times
F_N^2(\{q\}|\{k\})
\exp\left\{\rmi\sum_{j=1}^{N}\tau(q_j)
-\rmi\sum_{j=1}^{N+1}\tau(k_j)\right\}.
\end{multline}
Here the quantities $u_1,\dots,u_{N+1}$ are
\begin{equation}\label{pjs}
u_j=\frac{1}{{\sin}^2\alpha_j}+a,\qquad
j=1,\dots,N+1,
\end{equation}
where, for a later convenience, we introduced the notation
$a=4/L\cc$.

The sum \eqref{Gsum} is defined as the
sum over values of the set of integers $\{n\}=n_1,\dots,n_{N+1}$
and the integer $m$, which
label the solutions of the Bethe Ansatz equations,
\begin{equation}
\sum_{\{k\},\lambda}=\sum_{\substack{n_i\in\mathbb{Z}\\
n_1<\dots<n_{N+1}}}
\sum_{m=0}^N.
\end{equation}
Since every term of the sum in \eqref{Gsum},
being totally symmetric with respect to permutations of $k_j$'s,
vanishes as soon as any two
of $k_j$'s coincide, and since $k_i=k_j$ if $n_i=n_j$,
the sum over the ordered set of integers $\{n\}$ can be replaced by
independent sums over these integers,
\begin{equation}\label{sumsym}
\sum_{\substack{n_i\in\mathbb{Z}\\
n_1<\dots<n_{N+1}}}\longrightarrow
\frac{1}{(N+1)!} \sum_{n_1\in\mathbb{Z}}\dots \sum_{n_{N+1}\in\mathbb{Z}}.
\end{equation}
However, since the Bethe Ansatz equations for the momenta $\{k\}$ and the quasi-momentum
$\lambda$ are coupled, the $k_j$'s and $\lambda$ depend on all
these integers, $k_j=k_j(\{n\},m)$, $\lambda=\lambda(\{n\},m)$, so some further
transformations in \eqref{Gsum} are required to handle the summations.
Namely, it would be convenient that the sum over each individual $n_j$ implies
the summation over the admissible values of the momentum $k_j$ only; it would be
useful also to have an interpretation of the sum of $u_j^{-1}$'s standing
in the denominator in \eqref{Gsum}, which prevents a factorized
summation over admissible values of $k_j$'s in the multiple sum.

It turns out that all this can be achieved by considering the momenta $k_j$ as
functions of $\lambda$, namely, $k_j=k_j(\lambda)$,
in which $\lambda$ is allowed to take arbitrary (real) values;
the solutions of Bethe Ansatz equations \eqref{BAEkv} and \eqref{BAEvk}
correspond to the values $\lambda=\Lambda_m$,  $m=0,\dots,N$.
These values will follow as roots of certain
(transcendent) equation imposed on the set of functions $\{k(\lambda)\}$.
To be more precise, let us introduce a function
$z_n(\lambda)$, where $n\in\mathbb{Z}$ and $\lambda\in\mathbb{R}$,
as the solution of the equation
\begin{equation}\label{znl}
z_n (\lambda) =\pi n+\arccot\left(a z_n(\lambda)-\frac{2\lambda}{\cc}\right).
\end{equation}
As it can be easily seen (e.g., using graphical interpretation
by taking cotangent of both sides) that this equation, for every value of $n$, defines a
single-valued, continuous, monotonously increasing function of $\lambda$. The last
property can be seen by taking the derivative in $\lambda$ of \eqref{znl} and expressing
it in terms of $z(n;\lambda)$,
\begin{equation}\label{partialz}
\partial_\lambda  z_n(\lambda)
=\frac{2/\cc}{1+a+(az_n(\lambda)-2\lambda/ \cc)^2}.
\end{equation}
Now, given a set of integers $\{n\}=n_1,\dots,n_{N+1}$, we define the set of functions
$\{k(\lambda)\}=k_1(\lambda),\dots,k_{N+1}(\lambda)$ by setting
\begin{equation}\label{kjlambda}
k_j(\lambda)=\frac{2}{L} z_{n_j}(\lambda),\qquad j=1,\dots,N+1.
\end{equation}
Clearly, here each $k_j(\lambda)$ depends solely on $n_j$, as desired.
Furthermore, let us introduce functions
$\alpha_j(\lambda)$ and
$u_j(\lambda)$
by defining them by \eqref{deltajs} and \eqref{pjs}, respectively, where
$k_j$ is replaced by $k_j(\lambda)$. Namely, we set
$\alpha_j(\lambda):=\alpha(k_j(\lambda),\lambda)$ and
$u_j(\lambda):=u(k_j(\lambda),\lambda)$, where the functions 
$\alpha(k,\lambda)$ and $u(k,\lambda)$ are given by 
\begin{equation}\label{omegau}
\alpha(k,\lambda)=-\arccot\left(\frac{2(k-\lambda)}{\cc}\right)
\end{equation}
and
\begin{equation}\label{uau}
u(k,\lambda)=\frac{1}{{\sin}^2\alpha(k,\lambda)}+a=
1+a+\left(\frac{2(k-\lambda)}{\cc}\right)^2,
\end{equation}
respectively. Relation \eqref{partialz} then implies
\begin{equation}
\partial_\lambda
k_j(\lambda)
= \frac{a}{u_j(\lambda)}.
\end{equation}
Hence, introducing a $\lambda$-dependent ``total
momentum'' $P(\lambda)=\sum_{j}k_j(\lambda)$, we have
\begin{equation}\label{sumujs}
\sum_{j=1}^{N+1}u_j^{-1}(\lambda)=a^{-1}
\partial_\lambda
P(\lambda).
\end{equation}

The relation \eqref{sumujs} may be used to transform the expression 
$u_1^{-1}+\dots+u_N^{-1}$ from \eqref{Gsum}. For that we recall \eqref{sumdelta}, 
which implies
that the solutions of Bethe Ansatz equation in our treatment are restored at the
values $\lambda=\Lambda_0,\dots,\Lambda_N$, that is
\begin{equation}\label{sumalphas}
\sum_{j=1}^{N+1}\alpha_j(\Lambda_m)=-\pi m,
\qquad m=0,1,\dots,N.
\end{equation}
An equivalent way to impose this condition is to set
\begin{equation}\label{Plm}
P(\Lambda_m)=\frac{2\pi}{L}\sum_{j=1}^{N+1}n_j + \frac{2\pi}{L}m.
\end{equation}
Hence, for given values of the set of integers $\{n\}$ and integer $m$, the
sum of $u_j^{-1}$'s is exactly the derivative of the total momentum
at $\lambda=\Lambda_m$,
\begin{equation}\label{denom}
\sum_{j=1}^{N+1} u_j^{-1}=a^{-1} P'(\Lambda_m),
\end{equation}
where we used the notation
$P'(\Lambda_m)=\partial_\lambda P(\lambda)|_{\lambda=\Lambda_m}$.

To show how these considerations make it possible
to factorize of the summation in \eqref{Gsum}, let us
consider the sum over $m$ at some fixed set of values
of the integers $\{n\}$. Using \eqref{denom} for the denominator
and regarding the
remaining part (the numerator) of the summands
as some trial function $f(\lambda)$, we transform this sum
using the Dirac $\delta$-function as follows:
\begin{multline}
\sum_{m=0}^N \frac{f(\Lambda_m)}{P'(\Lambda_m)}
=\int_{-\infty}^{+\infty}\rmd\lambda\,
\sum_{m=0}^N\frac{\delta(\lambda-\Lambda_m)}{P'(\Lambda_m)}f(\lambda)
=\int_{-\infty}^{+\infty}\rmd\lambda\,
\sum_{m=0}^N\delta(P(\lambda)-P(\Lambda_m)) f(\lambda)
\\
=\frac{L}{2}\int_{-\infty}^{+\infty}\rmd\lambda\,
\sum_{m=0}^N\delta\left(\sum_{j=1}^{N+1}\alpha_j(\lambda)+\pi m\right) f(\lambda)
\\
=\frac{L}{4\pi}\int_{-\infty}^{+\infty}\rmd\lambda\int_{-\infty}^{+\infty}\rmd s\,
\frac{1-\rme^{\rmi(N+1)\pi s}}{1-\rme^{\rmi\pi s}}
\exp\left\{\rmi s\sum_{j=1}^{N+1}\alpha_j(\lambda)\right\}f(\lambda).
\end{multline}
Here we used relations \eqref{Plm} and \eqref{sumalphas} and at the last step we replaced the
Dirac $\delta$-function by its Fourier transform, to bring
the dependence on the $k_j(\lambda)$'s (defining
the $\alpha_j(\lambda)$'s) in a factorized form.

As a result, taking into account \eqref{sumsym},
we have the following representation for the impurity Green's 
function of the finite system
\begin{equation}\label{GdnN}
G_{\dn,N}(x,t|\{q\})
=\frac{a L}{4\pi}\int_{-\infty}^{+\infty}\rmd\lambda\int_{-\infty}^{+\infty}\rmd s\,
\frac{1-\rme^{\rmi(N+1)\pi s}}{1-\rme^{\rmi\pi s}}\,
\Xi_N(x,t|\{q\},\lambda;s),
\end{equation}
where
\begin{multline}\label{XiLong}
\Xi_N(x,t|\{q\},\lambda;s)=\frac{(N+1)!}{L^{2N+1}}
\sum_{k_1(\lambda)}\cdots\sum_{k_{N+1}(\lambda)}
\prod_{j=1}^{N+1}\frac{1}{u_j(\lambda)}
\\ \times
F_N^2(\{q\}|\{k\})
\exp\left\{\rmi\sum_{j=1}^{N}\tau(q_j)
-\rmi\sum_{j=1}^{N+1}\left[\tau(k_j(\lambda))-s\alpha_j(\lambda)\right]\right\}.
\end{multline}
Here each sum is performed over all the values of the
momentum $k_j(\lambda)$ defined in \eqref{kjlambda} as the corresponding
$n_j$ runs over all integer values.

The expression \eqref{XiLong} can be written in terms of
determinants of $N\times N$ matrices, using the procedure of
``insertion of the summation in the determinant'' similarly to
the infinite coupling case \cite{CIKT-93,IP-98}.
Using the total symmetry in permutation of the
momenta $k_1,\dots,k_{N+1}$ of the general term of the sum
in \eqref{XiLong}, one of the functions $F_N(\{q\}|\{k\})$,
when represented as the determinant \eqref{FN1},
can be replaced by the product
\begin{equation}
F_N(\{q\}|\{k\}) \longrightarrow 
(N+1)!\, 2^N\prod_{j=1}^{N} \frac{1}{q_j-k_j},
\end{equation}
while the second one, when using \eqref{FNN},
can be written as a sum of two terms
\begin{multline}
F_N(\{q\}|\{k\}) =
2^N\left[\det_{1\leq j,l\leq N}
\left(\frac{1}{q_j-k_l}-\frac{1}{q_j-k_{N+1}}\right)
-\det_{1\leq j,l\leq N}
\left(\frac{1}{q_j-k_l}\right)
\right]
\\
+2^N\det_{1\leq j,l\leq N}
\left(\frac{1}{q_j-k_l}\right).
\end{multline}
Since the matrix with the entries $1/(q_j-k_{N+1})$ has rank one,
the first term in this sum is a homogeneous linear function of its entries while
the second term is independent of this momentum.
This makes it possible to perform the summation with respect to
$k_{N+1}$. The summation with respect to the remaining momenta
$k_1,\dots,k_N$ are performed in the usual way.

As a result, the quantity $\Xi_N(x,t|\{q\},\lambda;s)$
is given in terms of determinants of
$N\times N$ matrices:
\begin{equation}\label{XiN}
\Xi_N(x,t|\{q\},\lambda;s)
=\det(S-R)+(G(x,t;\lambda;s)-1) \det S.
\end{equation}
The matrix $S=S(x,t|\{q\},\lambda;s)$ has entries
\begin{equation}\label{Sjl}
S_{jl}=
\frac{4\rme^{\rmi(\tau(q_j)+\tau(q_l))/2}}{L^2}\sum_{k(\lambda)}
\frac{\rme^{-\rmi\tau(k(\lambda))+\rmi s\alpha(k(\lambda),\lambda)}}
{u(k(\lambda),\lambda)(k(\lambda)-q_j)(k(\lambda)-q_l)}
\end{equation}
and the matrix $R=R(x,t|\{q\},\lambda;s)$, of rank one,
has entries
\begin{equation}\label{Rjl}
R_{jl}=
\frac{4\rme^{\rmi(\tau(q_j)+\tau(q_l))/2}}{L^3}
\sum_{k(\lambda)}
\frac{\rme^{-\rmi\tau(k(\lambda))+\rmi s\alpha(k(\lambda),\lambda)}}
{u(k(\lambda),\lambda)(k(\lambda)-q_j)}
\sum_{k(\lambda)}
\frac{\rme^{-\rmi\tau(k(\lambda))+\rmi s\alpha(k(\lambda),\lambda)}}
{u(k(\lambda),\lambda)(k(\lambda)-q_l)}.
\end{equation}
The function $G(x,t;\lambda;s)$ is
\begin{equation}
G(x,t;\lambda;s)=\frac{1}{L}\sum_{k(\lambda)}
\frac{\rme^{-\rmi\tau(k(\lambda))+\rmi s\alpha(k(\lambda),\lambda)}}
{u(k(\lambda),\lambda)}.
\end{equation}
The functions $\alpha(k,\lambda)$ and
$u(k,\lambda)$ are defined in \eqref{omegau} and \eqref{uau}, respectively; the
dependence on $x$ and $t$ is contained in the function $\tau(k)=t k^2-xk$.

\section{Results in the thermodynamic limit}

The thermodynamic limit is the limit in which $L,N\to\infty$ with the ratio
$N/L$ kept constant. A convenient property of the representation \eqref{GdnN} is that
in this limit it contains the Dirac $\delta$-function:
\begin{equation}
\frac{1-\rme^{\rmi(N+1)\pi s}}{1-\rme^{\rmi\pi s}}\to
2\delta(s),\qquad N\to\infty.
\end{equation}
Hence, in the deriving a thermodynamic limit expression for the Green's function
we can restrict ourselves in obtaining that of the
quantity $\Xi_N(x,t|\{q\},\lambda;s)\big|_{s=0}$.

To derive an thermodynamic limit expression for this quantity,
let us study the matrices $S=S(x,t|\{q\},\lambda;s)$
and $R=R(x,t|\{q\},\lambda;s)$ entering representation \eqref{XiN},
specifying everywhere $s=0$. We first consider the matrix $S$. It is
useful to consider separately its off-diagonal and diagonal entries.
For the off-diagonal entries, since $q_j\ne q_l$ for $j\ne l$,
we can use the relation
\begin{equation}
\frac{1}{k(\lambda)-q_j}
\frac{1}{k(\lambda)-q_l}=
\frac{1}{q_j-q_l}\left(\frac{1}{k(\lambda)-q_j}-\frac{1}{k(\lambda)-q_l}\right),
\end{equation}
and represent them in the form
\begin{equation}\label{Soffd}
S_{jl}=
\frac{2}{L}\rme^{\rmi(\tau(q_j)+\tau(q_l))/2}
\frac{E(q_j|\lambda)-E(q_l|\lambda)}{q_j-q_l},\qquad
j\ne l.
\end{equation}
Here the function $E(q|\lambda)$, $q\in\mathbb{R}$, is given by the expression
\begin{equation}\label{Eql}
E(q|\lambda)
=\frac{2}{L}\sum_{k(\lambda)}
\frac{1}{k(\lambda)-q}
\left(\frac{\rme^{-\rmi\tau(k(\lambda))}}{u(k(\lambda),\lambda)}
-\frac{\rme^{-\rmi\tau(q)}}{u(q,\lambda)}
\right)
+\frac{2(q-\lambda)\rme^{-\rmi\tau(q)}}{\cc u(q,\lambda)}.
\end{equation}
For $q=q_j$, where $q_j$ is one of the momenta in the
set $\{q\}$ of the sector $(N,0)$, it evaluates as
\begin{equation}\label{Eqjl}
E(q_j|\lambda)=\frac{2}{L}\sum_{k(\lambda)}
\frac{\rme^{-\rmi\tau(k(\lambda))}}{(k(\lambda)-q_j) u(k(\lambda),\lambda)},
\end{equation}
so \eqref{Soffd} reproduces \eqref{Sjl} for $j\ne l$.
Note, that the function $E(q|\lambda)$ is well-defined for arbitrary real values
of $q$; the expression in \eqref{Eql} follows upon extracting a formal singularity
at $k(\lambda)=q_j$ of the sum over $k(\lambda)$ in \eqref{Eqjl},
using a summation formula for $\sum_{k(\lambda)}1/(k(\lambda)-q_j)$,
see formula \eqref{sum1} of the appendix.
Similarly, for the diagonal entries of the matrix $S$ we obtain
\begin{multline}
S_{jj}
=\frac{4\rme^{-\rmi\tau(q_j)}}{L^2}
\sum_{k(\lambda)} \frac{1}{(k(\lambda)-q_j)^2}
\\  \times
\bigg\{
\frac{\rme^{-\rmi\tau(k(\lambda))}}
{u(k(\lambda),\lambda)}
-\frac{\rme^{-\rmi\tau(q_j)}}{u(q_j,\lambda)}
\left[1+
(k(\lambda)-q_j)
\left(-\rmi\tau'(q_j)-\frac{u'(q_j,\lambda)}{u(q_j,\lambda)}\right)
\right]
\bigg\}
\\
+\left[1+2a+\frac{4(q_j-\lambda)^2}{\cc^2}
+a(q_j-\lambda)
\left(-\rmi\tau'(q_j)-\frac{u'(q_j,\lambda)}{u(q_j,\lambda)}\right)
\right]
\frac{1}{u(q_j,\lambda)}
\\
=1+\frac{2}{L}\rme^{\rmi\tau(q_j)}
\partial_q E(q|\lambda)\big|_{q=q_j}.
\end{multline}
Here the first equality follows from
\eqref{Sjl} at $j=l$ by extracting a double-pole singularity
using a summation formula for $\sum_{k(\lambda)}1/(k(\lambda)-q)^2$,
see \eqref{sum2}, and the second equality follows just by
comparing the result with \eqref{Eql}. Introducing the function
\begin{equation}
V(q,q')=
\frac{2}{L}\rme^{\rmi(\tau(q)+\tau(q'))/2}
\frac{E(q|\lambda)-E(q'|\lambda)}{q-q'},\qquad q,q'\in\mathbb{R},
\end{equation}
we conclude that
\begin{equation}
S_{jl}=\delta_{jl}+V(q_j,q_l).
\end{equation}
In other words, $S=I+V$, where the matrix $V$ has entries $V_{jl}=V(q_j,q_l)$.

In similar manner, introducing the function
\begin{equation}
R(q,q')=\frac{1}{L}\rme^{\rmi(\tau(q)+\tau(q'))/2}E(q|\lambda)E(q'|\lambda),\qquad
q,q'\in\mathbb{R},
\end{equation}
for the entries of the matrix $R$ we have
\begin{equation}
R_{jl}=R(q_j,q_l).
\end{equation}
We note that, as a result, the matrices $V$ and $R$ are expressed in terms of the
functions $V(q,q')$ and $R(q,q')$, which are well-defined
for their arguments taking arbitrary real values. We also note that
the entries of the matrices $V$ and $R$ are all of order $1/L$, as $L$ is large.

Having all this in mind, we are now ready to address the problem of deriving the
thermodynamic limit of the quantity $\Xi_N(x,t|\{q\},\lambda;s)\big|_{s=0}$.
Consider the determinant of the matrix $S=I+V$; the case of the matrix
$S-R=I+V-R$ is similar. We have
\begin{equation}
\det(I+V)=\sum_{p=0}^{N} \frac{1}{p!} \sum_{q_{j_1}\in\{q\}} \dots
\sum_{q_{j_p}\in\{q\}}\det_{1\leq a,b \leq p} V(q_{j_a},q_{j_b}).
\end{equation}
Here each summation is performed over elements in the set $\{q\}$;
in the thermodynamic limit the summations turn into integrations.
To define the integrals, we need now specify the values of the momenta
in the set $\{q\}$. Recall that we are interested in the
Green's function of the ground state 
$\ket{\Omega_\Uparrow}=\ket{\Psi_{N,0}(\{q\})}$,
in which the momenta in the set $\{q\}$ satisfy the relations
$q_{j+1}-q_j=2\pi/L$, $q_1\geq -\kf$, $q_N\leq \kf$, where $\kf=\pi N/L$.
In the thermodynamic limit\ the quasi-momenta in the set $\{q\}$ are
uniformly distributed along the interval $[-\kf,\kf]$. Hence,
\begin{equation}
\frac{1}{L}\sum_{q_j\in\{q\}}\longrightarrow \frac{1}{2\pi}\int_{-\kf}^{\kf}\rmd q
\end{equation}
Correspondingly, let us introduce the function
\begin{equation}
\mathcal{V}(q,q') = \lim_{L,N\to\infty} \frac{L}{2\pi} V(q,q').
\end{equation}
Then, evaluating the limit, we obtain
\begin{equation}\label{Fdet}
\lim_{L,N\to\infty}
\det(I+V)=\sum_{p=0}^{\infty} \frac{1}{p!}
\int_{[-\kf,\kf]^p}\rmd q^p
\det_{1\leq a,b \leq p} \mathcal{V}(q_{j_a},q_{j_b})
=\det(1+\mathcal{V}).
\end{equation}
Here the second equality follows from the definition of the
Fredholm determinant of a linear integral operator $\mathcal{V}$;
the function $\mathcal{V}(q,q')$ is the kernel of the operator, which
is defined to act on functions in the interval $[-\kf,\kf]$,
\begin{equation}\label{kfkf}
(\mathcal{V}f)(q)=\int_{-\kf}^{\kf}\rmd q'\, \mathcal{V}(q,q') f(q').
\end{equation}
Similarly to \eqref{Fdet}, we have
\begin{equation}\label{Fdet2}
\lim_{L,N\to\infty}
\det(I+V-R)=\det(1+\mathcal{V}-\mathcal{R}),
\end{equation}
where the kernel of the operator $\mathcal{R}$ is defined by
\begin{equation}
\mathcal{R}(q,q') = \lim_{L,N\to\infty} \frac{L}{2\pi} R(q,q').
\end{equation}
Explicitly, the linear integral operator $\mathcal{V}$
entering \eqref{Fdet} and \eqref{Fdet2} possesses the kernel
\begin{equation}
\mathcal{V}(q,q')=
\frac{1}{\pi}\rme^{\rmi(\tau(q)+\tau(q'))/2}
\frac{e(q|\lambda)-e(q'|\lambda)}{q-q'},
\end{equation}
and the operator $\mathcal{R}$, entering \eqref{Fdet2}, possesses the kernel
\begin{equation}
\mathcal{R}(q,q')
=\frac{1}{2\pi}\rme^{\rmi(\tau(q)+\tau(q'))/2}e(q|\lambda)e(q'|\lambda).
\end{equation}
Here the function $e(q|\lambda)=e(q|x,t;\lambda)$ is the thermodynamic limit
of the function $E(q|\lambda)$,
\begin{equation}
e(q|\lambda)=\lim_{L\to\infty} E(q|\lambda).
\end{equation}
If we also define the function $g(x,t|\lambda)$ as the thermodynamic limit
of the function $G(x,t;\lambda;s)\big|_{s=0}$,
\begin{equation}
g(x,t;\lambda)=\lim_{L\to\infty} G(x,t;\lambda;0),
\end{equation}
then we have the following expression
for the quantity $\Xi_N(x,t|\{q\},\lambda;s)\big|_{s=0}$ in the thermodynamic limit:
\begin{equation}
\lim_{L,N\to\infty}\Xi_N(x,t|\{q\},\lambda;0)\big|_{s=0}=
\det(1+\mathcal{V}-\mathcal{R})+(g(x,t;\lambda)-1)\det(1+\mathcal{V}).
\end{equation}
We recall that the linear integral operators $\mathcal{V}$ and $\mathcal{R}$
depend on $x$, $t$, and $\lambda$
via the functions defining their kernels; they also depend implicitly on the Fermi
momentum $\kf$, defining the interval $[-\kf,\kf]$ where these operators act.

To finalize our calculation we have to derive the functions
$e(q|\lambda)$ and $g(x,t;\lambda)$. These functions contains summation over all
values of the momentum $k(\lambda)=2z_n(\lambda)/L$, where $z_n(\lambda)$
is the solution of \eqref{znl}, as $n$ runs over all integer values,
\begin{equation}
\sum_{k(\lambda)} f(k(\lambda))=\sum_{n\in\mathbb{Z}}f(2z_n(\lambda)/L).
\end{equation}
In the thermodynamic limit the sum turns into an integral. To obtain
the limit, we need to find a distribution of the values of the momentum
$k(\lambda)$. Denoting the value of $k(\lambda)$ for given $n$
as $[k(\lambda)]_n=2z_n(\lambda)/L$, we can define the density
\begin{equation}
\rho([k(\lambda)]_n)=\frac{1}{L([k(\lambda)]_{n+1}-[k(\lambda)]_{n})}.
\end{equation}
The sum turns into an integral by the rule
\begin{equation}
\frac{1}{L}\sum_{k(\lambda)}\longrightarrow
\int_{-\infty}^{+\infty}\rmd k\, \rho(k).
\end{equation}
Using that $z_{n+1}(\lambda)=z_{n}(\lambda)+\left(2\rho([k(\lambda)]_n)\right)^{-1}$
and recalling that $a=4/L\cc$, from \eqref{znl} we get
\begin{equation}
z_{n+1}(\lambda)-z_{n}(\lambda)=\pi-
\frac{1}{1+(a z_n(\lambda)-2\lambda/\cc)^2}\frac{a}{2\rho([k(\lambda)]_n)}+O(a^2).
\end{equation}
Replacing the left-hand side by $\left(2\rho([k(\lambda)]_n)\right)^{-1}$,
we obtain
\begin{equation}
2\pi\rho([k(\lambda)]_n)=1+\frac{a}{1+ 4([k(\lambda)]_n-\lambda)^2/\cc^2}+O(a^2).
\end{equation}
In the thermodynamic limit we have $a\to 0$, so we obtain $\rho(k)=(2\pi)^{-1}$.
Taking also into account the explicit expression
for the function $u(k,\lambda)$, see \eqref{omegau},
for the function $g(x,t;\lambda)$ we obtain
\begin{equation}\label{gxtl}
g(x,t;\lambda)=\frac{1}{2\pi} \int_{-\infty}^{+\infty}\rmd k\,
\frac{\rme^{-\rmi\tau(k)}}{1+4(k-\lambda)^2/\cc^2}.
\end{equation}
In turn, for the function $e(q|\lambda)=e(q|x,t;\lambda)$ we have
\begin{equation}\label{eql}
e(q|\lambda)=\frac{1}{\pi}\int_{-\infty}^{+\infty}\rmd k\,
\frac{\rme^{-\rmi\tau(k)}}{(k-q)(1+4(k-\lambda)^2/\cc^2)}
+\frac{2(q-\lambda)\rme^{-\rmi\tau(q)}}{\pi\cc(1+4(q-\lambda)^2/\cc^2)},
\end{equation}
where the integral has to be interpreted as a Cauchy principal
value.

We may now summarize our results about the Green's function
\eqref{Gdef}. Its 
is given in the thermodynamic limit in terms 
of an integral of Fredholm determinants:
\begin{equation}\label{mainresult}
G_\dn(x,t)=\frac{2}{\pi\cc}\int_{-\infty}^{+\infty}
\rmd\lambda\,
\left\{\det\big(1+\mathcal{V}-\mathcal{R}\big)
+(g(x,t;\lambda)-1)\det\big(1+\mathcal{V}\big)\right\}.
\end{equation}
The integral operators $\mathcal{V}=\mathcal{V}(x,t;\lambda)$ and
$\mathcal{R}=\mathcal{R}(x,t;\lambda)$ act on functions 
in the interval $[-\kf,\kf]$,
see \eqref{kfkf}, and possess kernels
\begin{equation}
\mathcal{V}(q,q')=
\frac{e_{+}(q|\lambda)e_{-}(q')-e_{-}(q)e_{+}(q'|\lambda)}{\pi(q-q')},\qquad
\mathcal{R}(q,q')=
\frac{e_{+}(q|\lambda)e_{+}(q'|\lambda)}{2\pi}.
\end{equation}
The functions defining the kernels are
\begin{equation}
e_{+}(q|\lambda)=e_{-}(q)\,e(q|\lambda),\qquad
e_{-}(q)=\rme^{\rmi \tau(q)/2}.
\end{equation}
The functions $g(x,t;\lambda)$ and $e(q|\lambda)=e(q|x,t;\lambda)$
are given by \eqref{gxtl} and \eqref{eql}, respectively, and
\begin{equation}
\tau(q)=tq^2-xq.	
\end{equation}

Representation \eqref{mainresult} is our main result. Let us discuss it in 
two special cases, namely, in the case of 
equal time, $t=0$, and 
in the limit of infinite coupling, $\cc\to\infty$.

At $t=0$ the anticommutation relations \eqref{anticomm}
give $G_\dn(x,t)=\delta(x)$. To show how this expression follows 
from \eqref{mainresult}, we first note that at $t=0$ the integrals 
in \eqref{gxtl} and \eqref{eql} 
can be explicitly evaluated and the results read
\begin{equation}
g(x,0;\lambda)
= \frac{\cc}{4} \rme^{\rmi x\lambda-\cc |x|/2},
\qquad
e(q|x,0;\lambda) 
= \cc\frac{\rme^{\rmi qx}-\rme^{\rmi x\lambda-\cc|x|/2}}
{2q-2\lambda-\rmi\cc\sgn(x)}.
\end{equation}
Now, we employ 
the definition of the Fredholm determinant as the expansion \eqref{Fdet}.
One can notice that after integration over $\lambda$ in \eqref{mainresult}
all terms with $p\ge 1$ in this expansion will vanish since 
all the singularities of the integrand lie in the lower
(respectively, upper) half-plane of the complex variable $\lambda$ 
for $x > 0$  ($x < 0$), while due to the
structure of the exponent $\rme^{\rmi\lambda x}$ the 
contour of integration can be shrunk in the upper (lower) $\lambda$ half-plane.
Finally, the $p=0$ term gives
\begin{equation}
G_\dn(x,0) = \frac{2}{\pi\cc} \int_{-\infty}^{+\infty}\rmd\lambda\, g(x,0;\lambda)
= \frac{\rme^{-\cc|x|/2}}{2\pi} \int_{-\infty}^{+\infty} \rmd\lambda\, 
\rme^{\rmi x\lambda}
= \delta(x),
\end{equation}
and the equal-time expression is reproduced.

Let us now consider the case of
the limit $\cc\to\infty$. Setting 
\begin{equation}
\lambda=-\frac{\cc}{2}\cot\vartheta
\end{equation}
and sending $\cc$ to infinity, we get
\begin{equation}
\lim_{\cc\to\infty}g(x,t;\lambda)={\sin}^2\vartheta\cdot g_\infty(x,t),\qquad
g_\infty(x,t)=\frac{1}{2\pi}\int_{-\infty}^{+\infty}\rmd k \rme^{-\rmi\tau(k)},
\end{equation}
and
\begin{equation}
\lim_{\cc\to\infty}e(q|\lambda)=
\frac{{\sin}^2\vartheta}{\pi}\int_{-\infty}^{+\infty}\rmd k\,
\frac{\rme^{-\rmi\tau(k)}}{k-q}
+\sin\vartheta\cos\vartheta\,
\rme^{-\rmi\tau(q)}=:e_\infty(q|\vartheta).
\end{equation}
Therefore, for the impurity Green's function in the $\cc\to\infty$ 
limit we obtain the following expression
\begin{equation}\label{Ginfty}
\lim_{c\to\infty}G_\dn(x,t)
=\frac{1}{\pi}\int_{0}^{\pi}\rmd\vartheta\,
\left\{\det\big(1+\mathcal{V}_\infty-\mathcal{R}_\infty\big)
+(g_\infty(x,t)-1)\det\big(1+\mathcal{V}_\infty\big)\right\},
\end{equation}
where the operators $\mathcal{V}_\infty=\mathcal{V}_\infty(x,t;\vartheta)$
and $\mathcal{R}_\infty=\mathcal{R}_\infty(x,t;\vartheta)$ possess kernels
\begin{equation}
\mathcal{V}_\infty(q,q')=\frac{\ell_{+}(q|\vartheta)\ell_{-}(q')
-\ell_{-}(q)\ell_{+}(q'|\vartheta) }{\pi(q-q')},\qquad
\mathcal{R}_\infty(q,q')=\frac{\ell_{+}(q|\vartheta)\ell_{+}(q'|\vartheta)}
{2\pi\,{\sin}^2\vartheta},
\end{equation}
and the functions $\ell_{+}(q|\vartheta)$ and $\ell_{-}(q)$ are
\begin{equation}
\ell_{+}(q|\vartheta)=\ell_{-}(q)\,e_\infty(q|\vartheta),\qquad \ell_{-}(q)=e_{-}(q).
\end{equation}
The representation \eqref{Ginfty} is the correlation function
$G_2^{(+)}(x,t;h,B)$ of paper \cite{IP-98} at
zero temperature in the case of a negative magnetic field $B$
(see \cite{IP-98}, Eqs.~(6.9)--(6.11)), evaluated at vanishing
chemical potential, $h=0$, and $B\to 0$. Hence, the known result
for the Green's function \eqref{Gdef} at $\cc=\infty$ is reproduced.

\section{Conclusion}

In this paper, we presented a closed-form analytic expression for 
the impurity Green's
function in a one-dimensional Fermi gas at zero temperature. A single
impurity interacts with gas particles through repulsive
$\delta$-function potential of arbitrary strength. This model is
exactly solvable and constitutes a particular case of the
model with Hamiltonian \eqref{Ham}, for which Bethe Ansatz eigenstates have the 
so-called ``nested'' form. 

Nested Bethe Ansatz wave functions are built for models
containing particles of more that one type (e.g., two 
types of fermions, which are often referred to as spin-up and
spin-down). These wave functions 
are known in a form very distinct from a single Slater 
determinant representation for the wave functions of the free Fermi
gas. However, when only one impurity (particle of other type) is added
to a free Fermi gas, any Bethe Ansatz wave function may be represented
as a single determinant. 

We discussed single determinant representation of the wave
functions in Sec. 2 of the paper. Clearly, such a representation
greatly simplifies calculations of form factors of local
operators, and, ultimately, correlation functions, as may be seen from
Sec. 3. There we represented Green's function \eqref{Gdef} as an infinite sum
of the determinants of $N\times N$ matrices, where $N$ is the number
of Fermi gas particles. In the thermodynamic limit of $N\to\infty$ at
a constant gas density the impurity Green's function is given as 
an integral of some Fredholm determinants, as discussed in Sec. 4.

The Fredholm determinant representation, \eqref{mainresult}, of the Green's
function \eqref{Gdef} is the main result of the present paper. This
representation extends our understanding of quantum interacting systems 
in one dimension in a number of ways. 

First, Fredholm determinant representations were so far known for
those systems with short-range interaction potential whose wave
functions vanish when any two particles approach each other. Wave
functions of a free Fermi gas is the basic example of such functions;
they all are often put under the name ``free-fermion-like wave
functions''. Wave function of the impurity model considered in the
present paper does not vanish as the impurity approaches any gas
particle. We thus found a Fredholm determinant representation for
Green's function in not-free-fermion-like model. 

Second, Fredholm
determinant may be viewed as just a special function in that it can be
calculated numerically and tabulated for any finite range of the
parameters entering its kernel. Therefore, the expression 
for the impurity Green's function given here can be used 
in comparisons of other methods against an exact result.   

Third, as far as the representation
for the Green's function involves Fredholm determinants of linear 
integral operators of the so-called integrable type,  
the standard technique of the analysis of such objects 
based on the matrix Riemann-Hilbert problem can be applied. 
This would make it possible for evaluation of asymptotics of 
the impurity Green's function at long time and large distance separations.

\section*{Acknowledgments}

The work of O.G. is supported by the European Research Council, Grant No. 279738-NEDFOQ.
The work of A.G.P. is partially supported by the Russian Science Foundation,
Grant No. 14-11-00598.

\appendix

\section{Summation formulas}

To transform the matrices $S$ and $R$, see \eqref{Sjl} and \eqref{Rjl}, 
in a form regular in the thermodynamic limit we used two summation formulas
in Sec.~4. Consider the function
\begin{equation}\label{gzl}
g(z,\lambda)=\sum_{n} \frac{1}{z-z_n(\lambda)},\qquad
z\in\mathbb{C},
\end{equation}
where the sum is taken over all solutions of \eqref{znl}.
It is assumed that the summation is organized
in a such way that the sum is convergent, thus
defining $g(z,\lambda)$ as a meromorphic function in $z$ with
simple poles at points $z_n(\lambda)$, $n\in\mathbb{Z}$, with the principal parts
$1/(z-z_n(\lambda))$ at these points.
The summation formulas used in Sec.~ 4 can be obtained  
from expressions for $g(z,\lambda)$ and $\partial_z g(z,\lambda)$
at $z=L q/2$, where $q$ is a solution of the Bethe Ansatz equations
in the sector $(N,0)$, given by \eqref{qj}.  

To derive an explicit form of the function $g(z,\lambda)$ 
it is useful to look at $z_n(\lambda)$'s as zeros
of some entire function (see, e.g., \cite{M-14}, Vol.~II, Chap.~10).
In our case this entire function can be easily inferred from
\eqref{znl} to be
\begin{equation}\label{entire}
f(z,\lambda)=\cos z -\left(a z- \frac{2\lambda}{\cc}\right)\sin z.
\end{equation}
Recalling that $a=4/L\cc$, we have the functional relation
\begin{equation}
f(z+\pi,\lambda)=-f(z,\lambda-2\pi/L).
\end{equation}
Correspondingly, zeros satisfy
\begin{equation}
z_{n+1}(\lambda)=\pi+z_{n}(\lambda-2\pi/L),
\end{equation}
that can also be seen directly from \eqref{znl}.
Writing $z_n(\lambda)=\pi n+z_0(\lambda-2\pi n/L)$, and taking into account
that $z_0(\lambda)\in[0,\pi]$, one can construct from
\eqref{znl} an estimate for $z_n(\lambda)$ as $n\to\pm\infty$. Namely,
we have
\begin{equation}
z_{n+1}(\lambda)-z_n(\lambda)=\pi + O(1/n),\qquad n\to\pm\infty.
\end{equation}
This estimate allows one to represent
the entire function $f(z,\lambda)$ as an infinite product.
Indeed, the estimate shows that the series $\sum_n 1/z_n^2(\lambda)$ converges, 
and function \eqref{entire} can be written as 
an infinite product, similarly to the well-known example  of the sine function.
By Borel's theorem for representation of an entire function as an infinite product 
we have the following formula
\begin{equation}\label{fprod}
f(z,\lambda)
=\rme^{2\lambda z/c}\prod_{n}\left(1-\frac{z}{z_n(\lambda)}\right)\rme^{z/z_n(\lambda)}.
\end{equation}
Comparing \eqref{gzl} and \eqref{fprod} we conclude (see also \cite{M-14}, Vol. II, Sec.~51, ex.~2) that
\begin{equation}\label{logdf}
g(z,\lambda)=\frac{\partial_z f(z,\lambda)}{f(z,\lambda)}.
\end{equation}
Substituting \eqref{entire} into \eqref{logdf}, we obtain that
the function $g(z,\lambda)$ has the following explicit form
\begin{equation}
g(z,\lambda)=-\frac{(1+a)\sin z+(a z-2\lambda/\cc)\cos z}
{\cos z-(a z- 2\lambda/\cc)\sin z}.
\end{equation}
Note that this expression is in agreement with the special case
$\lambda\to\pm\infty$, in which the zeros $z_n(\lambda)$ tend to those
of $\sin z$, and hence $g(z,\lambda)\to \cot z$.

Now we are ready to obtain the summation formulas used in Sec.~4. 
Setting $z=Lq_j/2$, where $q_j$ is a momentum belonging to the set
momenta $\{q\}$ of the sector $(N,0)$, we have
\begin{equation}
\frac{2}{L}\sum_{k(\lambda)}\frac{1}{k(\lambda)-q_j}=
\sum_{n}\frac{1}{z_n(\lambda)-Lq_j/2}=-g(z,\lambda)\big|_{z=Lq_j/2}.
\end{equation}
Since $q_j=2\pi m_j/L$, $m_j\in\mathbb{Z}$, we have
$\sin(Lq_j/2)=\sin \pi m_j=0$ and $\cos (Lq_j/2)=\cos\pi m_j=(-1)^{m_j}$, and we obtain
the first formula
\begin{equation}\label{sum1}
\frac{2}{L}\sum_{k(\lambda)}\frac{1}{k(\lambda)-q_j}
=\frac{2(q_j-\lambda)}{\cc},
\end{equation}
where we have used that $a=4/L\cc$.

The second formula, relevant for the diagonal entries of the matrix $S$,
follows from the derivative of the function $g(z,\lambda)$,
\begin{equation}
\frac{4}{L^2}\sum_{k(\lambda)}\frac{1}{(k(\lambda)-q_j)^2}
=-\partial_z g(z,\lambda)\big|_{z=Lq_j/2}.
\end{equation}
We have
\begin{equation}
\partial_z g(z,\lambda)=\frac{\partial_z^2f(z,\lambda) f(z,\lambda)-
\left(\partial_z f(z,\lambda)\right)^2}{f^2(z,\lambda)}
=-\frac{1 + 2 a + (a z-2\lambda/\cc)^2+ a^2 {\sin}^2 z}
{(\cos z-(a z- 2\lambda/\cc)\sin z)^2}.
\end{equation}
Hence,
\begin{equation}\label{sum2}
\frac{4}{L^2}\sum_{k(\lambda)}\frac{1}{(k(\lambda)-q_j)^2}
=1+2a+\frac{4(q_j-\lambda)^2}{\cc^2}.
\end{equation}

\bibliography{imp_bib}
\end{document}